\documentclass[12pt]{article}

\usepackage[totalheight = 23cm, totalwidth = 17cm]{geometry}
\usepackage{amssymb,amsmath,amsfonts,amsbsy,graphicx,bm,cite,color}

\definecolor{DarkRed}{rgb}{0.65,0,0}

\newcommand{\calC}{{\cal C}}
\newcommand{\calO}{{\cal O}}
\newcommand{\calP}{{\cal P}}

\begin{document}

\begin{titlepage}

\begin{center}

{\LARGE \bf 
Abundance of primordial black holes
\\
with local non-Gaussianity in peak theory
}

\vskip 1.0cm

{\large
Chul-Moon Yoo$^{a}$,
Jinn-Ouk Gong$^{b}$ 
and Shuichiro Yokoyama$^{c,d}$ 
}

\vskip 0.5cm

{\it
$^{a}$Division of Particle and Astrophysical Science, Graduate School of Science
\\
Nagoya University, Nagoya 464-8602, Japan
\\
$^{b}$Korea Astronomy and Space Science Institute, Daejeon 34055, Korea
\\
$^{c}$Kobayashi-Maskawa Institute, Nagoya University, Nagoya 464-8602, Japan
\\ 
$^{d}$Kavli Institute for the Physics and Mathematics of the Universe, The University of Tokyo
\\
Kashiwa, Chiba 277-8583, Japan
}

\vskip 1.2cm

\end{center}

\begin{abstract}

We discuss the effect of local type non-Gaussianity on the abundance of primordial black holes (PBH)
based on the peak theory.
We provide the PBH formation criterion based on the so-called compaction function
and use the peak theory statistics associated with the curvature perturbation 
with the local type non-Gaussianity.
Providing a method to estimate the PBH abundance, we demonstrate the effects of non-Gaussianity.
It is explicitly shown that the value of non-linear parameter $|f_{\rm NL}| \sim 1$ 
induces a similar effect to a few factors of difference in the amplitude of the power spectrum.

\end{abstract}

\end{titlepage}

\newpage

\section{Introduction}
\label{sec:intro}

Primordial black holes (PBHs) repeatedly come up as a hot topic in cosmology and astrophysics since the pioneering works of Zel'dovich and Novikov~\cite{1967SvA....10..602Z} and Hawking~\cite{Hawking:1971ei}. For instance, PBHs are still fascinating candidate for dark matter (e.g., see~\cite{1975Natur.253..251C,Carr:1975qj,GarciaBellido:1996qt,Jedamzik:1999am,Frampton:2010sw,Kawasaki:2012wr,Kohri:2012yw,Carr:2016drx,Axelrod:2016nkp,Ezquiaga:2017fvi,Clesse:2017bsw,Kohri:2018qtx} and references therein), and observational constraints are frequently updated (see e.g.~\cite{Carr:2009jm,Carr:2016drx}). Number of interesting scenarios which produce substantial number of PBHs are proposed (see e.g. a recent review~\cite{Sasaki:2018dmp} and references therein).

One of inevitable issues is the estimation of PBH abundance. Simplest conventional way to estimate PBH abundance is the Press-Schechter (PS) formalism~\cite{Press:1973iz}, where the Gaussian distribution of a perturbation variable and its threshold for PBH formation are assumed. The validity of these assumptions has been under discussion for a long time (see e.g.~\cite{Carr:1975qj,1978SvA....22..129N,1980SvA....24..147N,Shibata:1999zs,Niemeyer:1999ak,Musco:2004ak,Polnarev:2006aa,Musco:2008hv,Musco:2012au,Polnarev:2012bi,Harada:2013epa,Nakama:2013ica,Young:2014ana,Nakama:2014fra,Harada:2015yda,Musco:2018rwt}). Recently, a more plausible procedure for PBH formation in a radiation dominated era is proposed in~\cite{Yoo:2018esr}, where an enhanced feature is assumed in the primordial power spectrum around some specific scale which enters the horizon in a radiation dominated era. Another important assumption is the random Gaussian distribution of the curvature perturbation. Then, a reliable procedure for the estimation of PBH abundance is derived based on the peak theory~\cite{doroshkevich,Bardeen:1985tr} of the curvature perturbation, by taking the non-linear effect between the curvature perturbation and the density perturbation into account (see e.g.~\cite{Kawasaki:2019mbl,DeLuca:2019qsy,Young:2019yug} for the significance of the non-linearity).

The purpose of this article is to show the impact of the local type non-Gaussianity of the curvature perturbation extending the procedure given in~\cite{Yoo:2018esr}. 
We note that in~\cite{Yoo:2018esr}, 
the intrinsic non-Gaussianity that originates from the non-linear relation between the curvature perturbation and the density perturbation is taken into account. 
However, throughout~\cite{Yoo:2018esr} the curvature perturbation is assumed to be a random Gaussian variable. 
In this article, we provide a method to introduce the local type 
non-Gaussianity of the curvature perturbation into the procedure provided in~\cite{Yoo:2018esr}.  
The effect of non-Gaussianity on the PBH abundance and the spatial distribution of PBHs have been discussed in~\cite{Bullock:1996at, Ivanov:1997ia, Yokoyama:1998pt, Hidalgo:2007vk, Saito:2008em, Bugaev:2011wy, Byrnes:2012yx, Linde:2012bt, Bugaev:2013vba, Young:2013oia, Tada:2015noa, Young:2015kda, Young:2015cyn, Ando:2017veq, Franciolini:2018vbk, Ezquiaga:2018gbw, Ando:2018nge, Atal:2018neu, Passaglia:2018ixg}. However, since a new plausible procedure is proposed, it is important to check how the previously reported effect of the non-Gaussianity works in the new procedure. Providing a method to estimate PBH abundance including the local type non-Gaussianity, we demonstrate the effect of the non-Gaussianity of the curvature perturbation.

This article is organized as follows. In Section~\ref{sec:profile}, the typical profile of the curvature perturbation is derived for a given scale and amplitude of the peak with the non-linear parameter $f_{\rm NL}$. The general implicit expression for the PBH fraction is derived in Section~\ref{sec:fraction}. Two specific primordial power spectra are shown as examples to present the impact of non-Gaussianity in Section~\ref{sec:examples}. Section~\ref{sec:summary} is devoted to a short summary. Some technical detail is given in the appendices. Throughout this article, we use the geometrized units in which both the speed of light and the Newton's gravitational constant are unity, $c=G=1$.

\section{Typical peak profile with local non-Gaussianity}
\label{sec:profile}

In this article, we consider the spatial metric given by
\begin{equation}
ds_3^2 = a^2 e^{-2\zeta} \tilde\gamma_{ij} dx^i dx^j 
\, ,
\label{eq:zetadef}
\end{equation}
and now we assume that the reference spatial metric $\tilde{\gamma}_{ij}$ is flat, and that the curvature perturbation $\zeta$ is given by the following form with the local type non-Gaussianity:
\begin{equation}
\zeta = \zeta_{\rm G} - \frac35 f_{\rm NL} \left( \zeta_{\rm G}^2 
- \left\langle \zeta_{\rm G}^2 \right\rangle \right)
\, , 
\label{eq:nonGzeta}
\end{equation}
where the brackets denote the ensemble average, and $\zeta_{\rm G}$ is the random Gaussian distribution with the power spectrum $\calP(k)$ defined by 
\begin{equation}
\left\langle \tilde\zeta_{\rm G}^*(\bm k) \tilde\zeta_{\rm G}(\bm k') \right\rangle
=
(2\pi)^3\delta(\bm k-\bm k') \frac{2\pi^2}{k^3}\calP(k)
\, , 
\end{equation}
where $\tilde\zeta_{\rm G}(\bm k)$ is the Fourier transform of $\zeta_{\rm G}$.
Note that we follow the same sign convention of $\zeta$ as~\cite{Lyth:2004gb,Yoo:2018esr}, so the negative sign of the non-linear term in \eqref{eq:nonGzeta} keeps $f_{\rm NL}$ consistent with the conventional definition, e.g. adopted in the Planck 2018 report on non-Gaussianity~\cite{Akrami:2019izv}.

Let us focus on a high peak of $\zeta_{\rm G}$. Here we assume that all peaks of $\zeta_{\rm G}$ are peaks of $\zeta$. The validity of this assumption is discussed in Appendix~\ref{app:peaks}. According to the peak theory~\cite{Bardeen:1985tr}, for a high peak at $\bm r = \bm 0$, we obtain the probability distribution for the spherical profile of $\zeta_{\rm G}(r)$ as follows:
\begin{equation}
P(\zeta_{\rm G}(r)|\mu,k_*)
=
\frac{1}{\sqrt{2\pi}\sigma_\zeta} \exp \left\{ -\frac{1}{2\sigma_\zeta^2}
\left[ \zeta_{\rm G}(r) - \bar\zeta_{\rm G}(r) \right]^2 \right\}
\, , 
\label{eq:PDF}
\end{equation}
where $\mu$ and $k_*$ are respectively the amplitude and the curvature scale of the Gaussian peak:
\begin{align}
\mu 
& =
-\left. \zeta_{\rm G} \right|_{r=0}
\, , 
\label{eq:mudef}
\\
k_*^2
& =
\frac{\left. \triangle\zeta_{\rm G} \right|_{r=0}}{\mu}
\, . 
\label{eq:ksdef}
\end{align}
The mean $\bar{\zeta}_{\rm G} (r)$ and the variance $\sigma_\zeta$ are respectively given as
\begin{align}
-\frac{\bar\zeta_{\rm G}(r)}{\sigma_0}
& =
\frac{\mu/\sigma_0}{1-\gamma^2} \left( \psi + \frac{1}{3} R_*^2 \triangle\psi \right)
-\frac{\mu k_*^2 /\sigma_2}{\gamma(1-\gamma^2)} 
\left( \gamma^2\psi + \frac{1}{3} R_*^2 \triangle\psi \right)
\, , 
\label{eq:zetabar}
\\
\frac{\sigma_\zeta^2}{\sigma_0^2}
& =
1-\frac{1}{1-\gamma^2} \psi^2
-\frac{1}{\gamma^2(1-\gamma^2)} \left( 2\gamma^2\psi + \frac{1}{3} R_*^2 \triangle\psi \right)
\frac{1}{3} R_*^2 \triangle\psi
\nonumber\\
&
\quad
- \frac{5}{\gamma^2} \left( \frac{\psi'}{r} - \frac{1}{3} \triangle\psi \right)^2 R_*^2
- \frac{1}{\gamma^2} R_*^2 {\psi'}^2
\, ,
\end{align}
with $\gamma = \sigma_1^2/(\sigma_0\sigma_2)$, $R_* = \sqrt{3}\sigma_1/\sigma_2$ and 
\begin{equation}
\psi(r)
=
\frac{1}{\sigma_0^2} \int\frac{dk}{k} \frac{\sin(kr)}{kr} \calP(k)
\, . 
\label{eq:psi}
\end{equation}
In the above, $\sigma_n$ is a $n$-th order gradient moment for $\zeta_{\rm G}$ defined by 
\begin{equation}
\sigma_n^2 \equiv \int \frac{dk}{k} k^{2n} \calP(k)
\, . 
\end{equation}
As in \eqref{eq:nonGzeta}, here, the non-Gaussian curvature perturbation $\zeta$ is explicitly given as a function of $\zeta_{\rm G}$. Then, by using \eqref{eq:PDF}, we can derive the typical profile $\bar\zeta(r;\mu,k_*)$ for $\zeta(r)$ as
\begin{align}
-\bar\zeta(r;\mu,k_*)
& =
-\int \zeta P(\zeta_{\rm G}|\mu,k_*) d\zeta_{\rm G}
\cr
& =
-\int \left[ \zeta_{\rm G} - \frac35 f_{\rm NL} 
\left( \zeta_{\rm G}^2 - \left\langle \zeta_{\rm G}^2 \right\rangle \right) \right]
P(\zeta_{\rm G}|\mu,k_*) d\zeta_{\rm G}
\cr
& =
-\bar\zeta_{\rm G} + \frac35 f_{\rm NL} \left(
{\bar\zeta_{\rm G}}^2 + \sigma_\zeta^2 -\sigma_0^2 \right)
\, . 
\label{eq:typical}
\end{align}

\section{General expression for the PBH fraction}
\label{sec:fraction}

Hereafter the calculations are mostly implicit and can be numerically done in general, and the basic procedure of the calculations are shown in \cite{Yoo:2018esr}. Thus, we just give a sketch of the implicit calculations with some details being given in Appendix~\ref{app:criteria}).

We consider that the criterion for PBH formation is given in terms of the so-called compaction function $\calC$, which is defined as the mass excess at a certain region. As shown Appendix~\ref{app:criteria}, $\calC$ can be written as a function of $\zeta$. PBH would be formed when the maximum value of the compaction function, $\calC^{\rm max}$, is larger than a threshold $\calC_{\rm th}$. Evaluating \eqref{eq:forrm} with respect to the typical profile $\bar\zeta$, we may obtain the value of $r_{\rm m}$, at which the compaction function takes the maximum value ${\mathcal C}^{\rm max}$, as a function of $\mu$ and $k_*$. That is, $r_{\rm m} = \bar r_{\rm m}(\mu,k_*)$ where an overbar means it is evaluated by using $\bar\zeta$, which is a function of $\mu$ and $k_*$. Then, $\calC^{\rm max}$ can be also expressed as a function of $\mu$ and $k_*$ as $\calC^{\rm max} = \bar\calC^{\rm max}(\mu,k_*)$. Note that $\bar{r}_{\rm m}$ and $\bar\calC^{\rm max}$ depend on the non-linear parameter $f_{\rm NL}$ through the non-Gaussian correction in the typical profile $\bar{\zeta}$ as shown in \eqref{eq:typical}. By identifying $\bar\calC^{\rm max}(\mu,k_*)$ as the threshold value $\calC_{\rm th} \approx 0.267$, we define the threshold $\mu_{\rm th}$ as a function of $k_*$:
\begin{equation}
\mu_{\rm th} = \mu_{\rm th}(k_*)
\, . 
\end{equation}
We may also regard the PBH mass $M$, which is defined by $r_{\rm m}$ [see \eqref{eq:kM} in Appendix \ref{app:criteria}], as a function of $\mu$ and $k_*$ by identifying $r_{\rm m} = \bar r_{\rm m}(\mu,k_*)$ and $\zeta = \bar\zeta(\bar r_{\rm m};\mu,k_*)$:
\begin{equation}
\label{eq:Mmuk}
M = \bar M(\mu,k_*)
\, . 
\end{equation}
Then, eliminating $k_*$ from the above two equations, the threshold value of $\mu_{\rm th}$ is given as a function of the PBH mass $M$:
\begin{equation}
\mu_{\rm th} = \mu_{\rm th}(M)
\, ,
\end{equation}
and this would be modified by the non-Gaussian effect. As will be shown later, in particular, for the case with the extended power spectrum we may have a minimum value of $\mu$ for a fixed value of $M$. Let  the function $\mu_{\rm min}(M)$ denote this minimum value. Then, the relevant region of $\mu$ for PBH formation with the mass $M$ is given by 
\begin{equation}
\mu > \mu_{\rm b} \equiv \max \left\{ \mu_{\rm min}(M), \mu_{\rm th}^{(M)}(M) \right\}
\, . 
\label{eq:mub}
\end{equation}

Since the parameters $\mu$ and $k_*$ are based on the Gaussian random variable, we can use the standard expression for the peak number density: 
\begin{equation}
n_{\rm pk}(\mu, k_*) d\mu dk_*
=
\frac{2 }{3^{3/2}(2\pi)^{3/2}} \mu k_*\frac{\sigma_2^2}{\sigma_0\sigma_1^3}
f \left( \frac{\mu k_*^2}{\sigma_2} \right)
P_1 \left( \frac{\mu}{\sigma_0},\frac{\mu k_*^2}{\sigma_2} \right) 
d\mu dk_*
\, ,  
\label{eq:nks}
\end{equation}
where the function $f$ and the probability $P_1$ are given by
\begin{align}
f(x)
& =
\frac{1}{2}x(x^2-3)
\left[ {\rm erf} \left( \frac{1}{2} \sqrt{\frac{5}{2}} x \right)
+ {\rm erf} \left( \sqrt{\frac{5}{2}} x \right) \right]
\cr
&
\quad
+ \sqrt{\frac{2}{5\pi}} \left[
\left( \frac{8}{5} + \frac{31}{4} x^2 \right) e^{-5x^2/8}
+ \left( -\frac{8}{5} + \frac{1}{2} x^2 \right) e^{-5x^2/2}
\right]
\, , 
\label{eq:funcf}
\\
P_1 \left( \frac{\mu}{\sigma_0},\frac{\mu k_*^2}{\sigma_2} \right) 
& =
\frac{1}{2\pi\sqrt{1-\gamma^2}}
\exp \left\{
-\frac{\mu^2}{2} \left[
\frac{1}{\sigma_0^2} + \frac{1}{\sigma_2^2(1-\gamma^2)} 
\left( k_*^2 - \frac{\sigma_1^2}{\sigma_0^2} \right)^2
\right]
\right\}
\, .
\label{eq:p1}
\end{align}
Further, since the direct observable is not $k_*$ but the PBH mass $M$, we change the variable from $k_*$ to $M$ as follows:
\begin{align}
n_{\rm pk}(\mu, M) d\mu dM
& =
3^{-3/2}(2\pi)^{-3/2}
\frac{\sigma_2^2}{\sigma_0\sigma_1^3}
\mu k_*
f \left( \frac{\mu k_*^2}{\sigma_2} \right)
P_1 \left( \frac{\mu}{\sigma_0},\frac{\mu k_*^2}{\sigma_2} \right) 
\cr
&
\quad
\times \left| \left[ \frac{1}{\bar r_{\rm m}} - \bar\zeta'(\bar r_{\rm m}) \right] \frac{d\bar r_{\rm m}}{dk_*}
- \left( \frac{d\bar\zeta}{dk_*} \right)_{r=\bar r_{\rm m}} \right|^{-1}
d\mu d\log M
\, ,  
\end{align}
where $k_*$ should be regarded as a function of $\mu$ and $M$ as in \eqref{eq:Mmuk}.
%
%
%
%
%
%
%
%
By counting the number of peaks whose $\mu$ is larger than the threshold value $\mu_{\rm b}$, we can evaluate the number density of PBHs as 
\begin{equation}
n_{\rm PBH} d\log M
=
\left[ \int^\infty_{\mu_{\rm b}} d\mu~n_{\rm pk}(\mu,M) \right] M d \log M
\, . 
\end{equation}
We also note that the scale factor $a$ is a function of $M$ as shown in \eqref{eq:kM}. Then, the fraction of PBHs to the total density $\beta_0 d\log M$ can be given by 
\begin{align}
\beta_0 d\log M
& =
\frac{M n_{\rm PBH}}{\rho a^3} d\log M
\cr
& =
\frac{4\pi}{3} n_{\rm PBH} k_{\rm eq}^{-3}
\left( \frac{M}{M_{\rm eq}} \right)^{3/2} d\log M
\cr
& =
\frac{2k_{\rm eq}^{-3}}{3^{5/2} (2\pi)^{1/2}}
\frac{\sigma_2^2}{\sigma_0\sigma_1^3}
\left( \frac{M}{M_{\rm eq}} \right)^{3/2}
\Bigg\{
\int^\infty_{\mu_{\rm b}} d\mu 
\mu k_* f \left( \frac{\mu k_*^2}{\sigma_2} \right)
P_1 \left( \frac{\mu}{\sigma_0},\frac{\mu k_*^2}{\sigma_2} \right)
\cr
&
\hspace{11em}
\times
\left| \left[ \frac{1}{\bar r_{\rm m}} - \bar \zeta'(\bar r_{\rm m}) \right] 
\frac{d\bar r_{\rm m}}{dk_*}
- \left( \frac{d \bar \zeta}{d k_*} \right)_{r=\bar r_{\rm m}} \right|^{-1}
\Bigg\} d\log M
\, , 
\label{eq:beta_general}
\end{align}
where $M_{\rm eq}$ and $k_{\rm eq}$ are respectively corresponding to the formed PBH mass and the comoving wave number reentering the horizon, at the matter radiation equality.

\section{Effect of local non-Gaussianity on PBH abundance}
\label{sec:examples}

Given the general expression for the PBH abundance \eqref{eq:beta_general}, now we employ explicit examples with specific power spectra and investigate the effect of the local non-Gaussianity on the abundance of PBHs.

\subsection{Monochromatic power spectrum}
\label{sec:monochrom}

\begin{figure}[htbp]
\begin{center}
\includegraphics[width=0.3\textwidth]{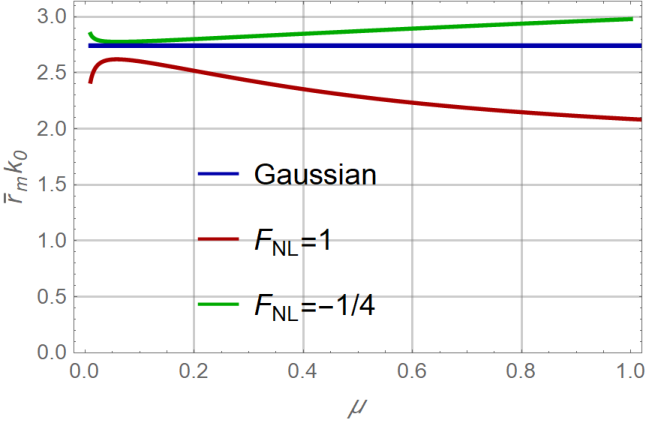}
\hspace{0.03\textwidth}
\includegraphics[width=0.3\textwidth]{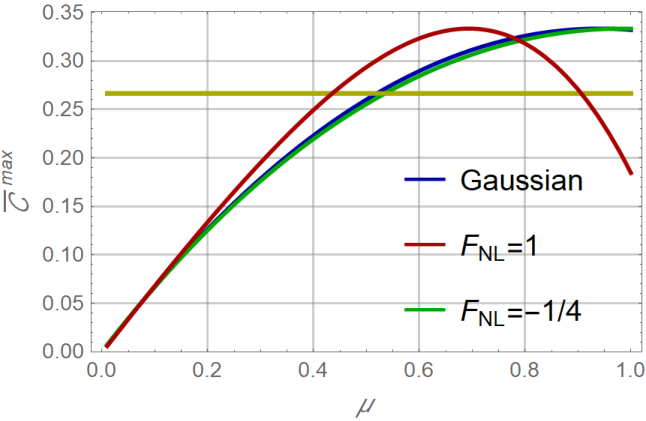}
\hspace{0.03\textwidth}
\includegraphics[width=0.3\textwidth]{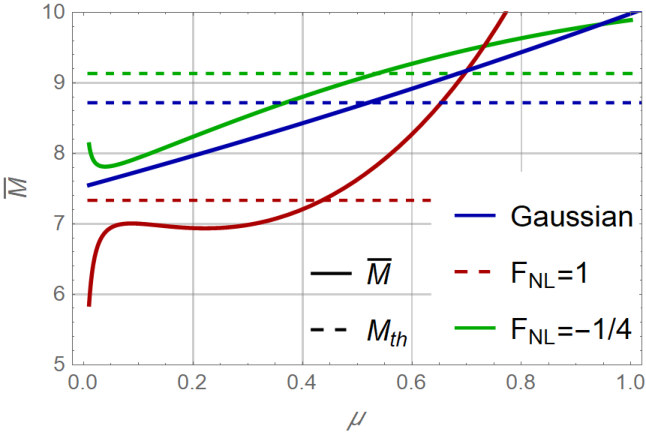}
\end{center}
\caption{(Left panel) $\bar r_{\rm m}(\mu)$, (middle panel) $\bar\calC^{\rm max}(\mu)$ and (right panel) $\bar M(\mu)$ for $\sigma_0=0.06$. Here, $F_{\rm NL} \equiv 3f_{\rm NL}/5$ for notational simplicity.}
\label{fig:mono}
\end{figure}

Let us consider the monochromatic power spectrum given by 
\begin{equation}
\calP(k) = \sigma_0^2 k_0 \delta(k-k_0)
\, .
\label{eq:monopower}
\end{equation}
For this case, gradient moments are explicitly calculated as $\sigma_n^2=\sigma_0^2k_0^{2n}$. Then, replacing $k_*$ by $k_0$ in \eqref{eq:zetabar}, we find
\begin{equation}
-\bar\zeta_{\rm G} = \mu\psi(r) = \mu\frac{\sin(k_0r)}{k_0r}
\, ,  
\end{equation}
where $\psi$ is calculated from \eqref{eq:psi}. Since the value of $k_*$ is fixed to be $k_* = k_0$, we do not have $k_*$ dependence for $\bar r_{\rm m}$, $\bar \calC^{\rm max}$, $\mu_{\rm th}$ and $\bar M$. The value of $\mu_{\rm th}$ is converted into the threshold value of $M_{\rm th}$ through the function $\bar M(\mu)$, i.e. $M_{\rm th}=\bar M(\mu_{\rm th})$. Through the non-Gaussian correction in the typical profile $\bar{\zeta}$, as mentioned in the previous section, $\bar r_{\rm m}$, $\bar\calC^{\rm max}$ and also $\bar M$ are dependent on $f_{\rm NL}$. Figure~\ref{fig:mono} shows the non-Gaussian effects on $\bar r_{\rm m}$, $\bar\calC^{\rm max}$ and $\bar M$ as functions of the peak amplitude $\mu$. Here, we take $\sigma_0=0.06$ and in each panel the blue, red, and green lines are respectively for the Gaussian case, i.e. $f_{\rm NL} = 0$, $3f_{\rm NL}/5 = 1$ and $3 f_{\rm NL}/5 = -1/4$. In the panel for $\bar\calC^{\rm max}$, the yellow horizontal line represents the threshold for PBH formation. The dashed lines in the panel for $\bar{M}$ are also the thresholds with respect to $M$.

\begin{figure}[htbp]
\begin{center}
\includegraphics[width=0.5\textwidth]{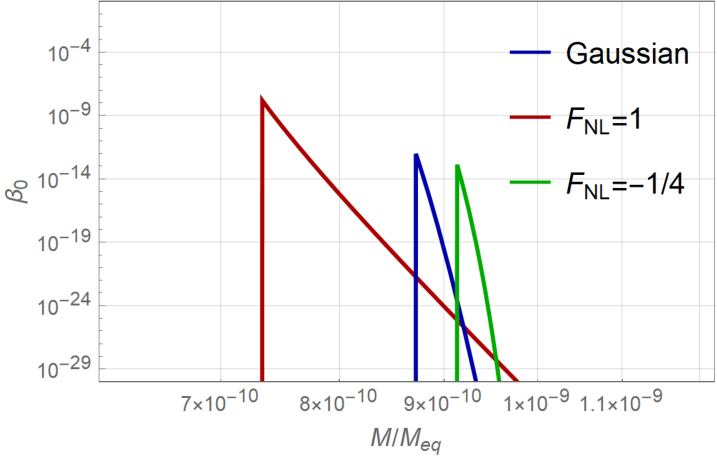}
\caption{
\baselineskip5.5mm
$\beta_0$ as a function of $M$ with $\sigma_0=0.06$ and  $k_0=10^5k_{\rm eq}$. The PBH fraction increases with increasing the value of $f_{\rm NL}$. 
}
\label{fig:beta_mono}
\end{center}
\end{figure}

Since the probability distribution $P_1$ is given by 
\begin{equation}
P_1 \left( \frac{\mu}{\sigma_0}, \frac{\mu k_*^2}{\sigma_2} \right) 
\underset{\gamma\rightarrow1}{\longrightarrow}
\frac{1}{2\sqrt{2\pi}} \frac{\sigma_2}{\mu k_*} \delta(k_*-k_0)
\exp \left( -\frac{\mu^2}{2\sigma_0^2} \right)
\, , 
\end{equation}
the peak number density can be written as 
\begin{equation}
n_{\rm pk}(\mu) d\mu
=
3^{-3/2}(2\pi)^{-2} \frac{1}{\sigma_0} k_0^3 f\left( \frac{\mu}{\sigma_0} \right)
\exp\left( -\frac{\mu^2}{2\sigma_0^2} \right) d\mu
\, . 
\end{equation}
Changing the variable from $\mu$ to $M$ through the function $M=\bar M(\mu)$, and taking the threshold value $M_{\rm th}$ into account, we obtain the following expression for the PBH number density:
\begin{equation}
n_{\rm PBH} d\log M
=
3^{-3/2}(2\pi)^{-2} \frac{1}{\sigma_0} k_0^3 f\left( \frac{\mu}{\sigma_0} \right)
\exp\left( -\frac{\mu^2}{2\sigma_0^2} \right) M \frac{d\mu}{dM} \Theta(M-M_{\rm th}) dM
\, , 
\end{equation}
where $\mu$ should be regarded as a function of $M$ through the relation $M=\bar M(\mu)$. Finally, we obtain 
\begin{equation}
\beta_0 d\log M
=
\frac{1}{3^{5/2}\pi\sigma_0}
\left( \frac{M}{M_{\rm eq}} \right)^{3/2}
\left( \frac{k_0}{k_{\rm eq}} \right)^3
f\left( \frac{\mu}{\sigma_0} \right)
\exp \left( -\frac{\mu^2}{2\sigma_0^2} \right) M\frac{d\mu}{dM} \Theta(M-M_{\rm th}) dM
\, . 
\end{equation}
Note that the functional form of this expression is completely the same as (80) in \cite{Yoo:2018esr}. The non-Gaussian correction in $\beta_0$ appears just through the modifications of $M_{\rm th}$ and the value of $\mu$ at $M_{\rm th}$. The PBH fraction $\beta_0$ for $3f_{\rm NL}/5=-1/4$, $0$ and $1$ is depicted as a function of $M$ with $\sigma_0=0.06$ and $k_0=10^5k_{\rm eq}$ in Figure~\ref{fig:beta_mono}. As can be seen, for larger $f_{\rm NL}$ the cut-off scale for PBH mass is shifted to smaller value. This is due to the smaller $M_{\rm th}$ for larger $f_{\rm NL}$ as shown in the right panel in Figure~\ref{fig:mono}. Furthermore, the peak value of the PBH fraction $\beta_0$ increases with increasing the value of $f_{\rm NL}$. This is due to the fact that the value of $ \mu$ at $M=M_{\rm th}$ becomes smaller for larger value of $f_{\rm NL}$, as can be seen in the right panel in Figure~\ref{fig:mono}.

\subsection{Extended power spectrum}
\label{sec:extended}

\begin{figure}[htb]
\begin{center}
\includegraphics[width=0.3\textwidth]{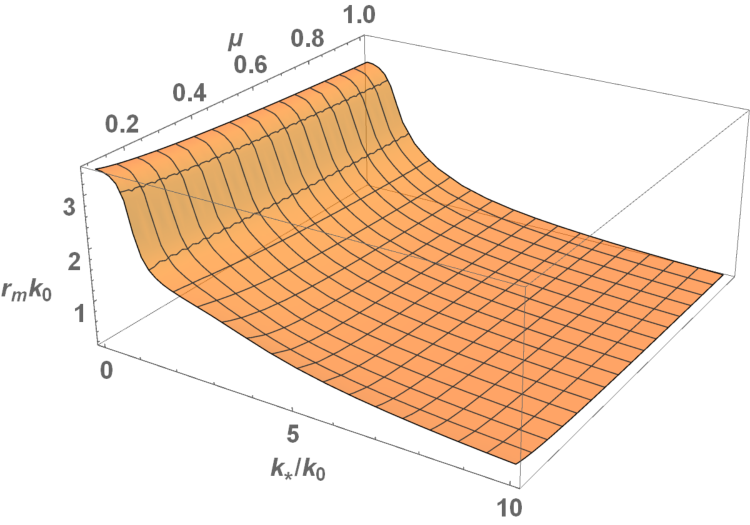}
\hspace{0.03\textwidth}
\includegraphics[width=0.3\textwidth]{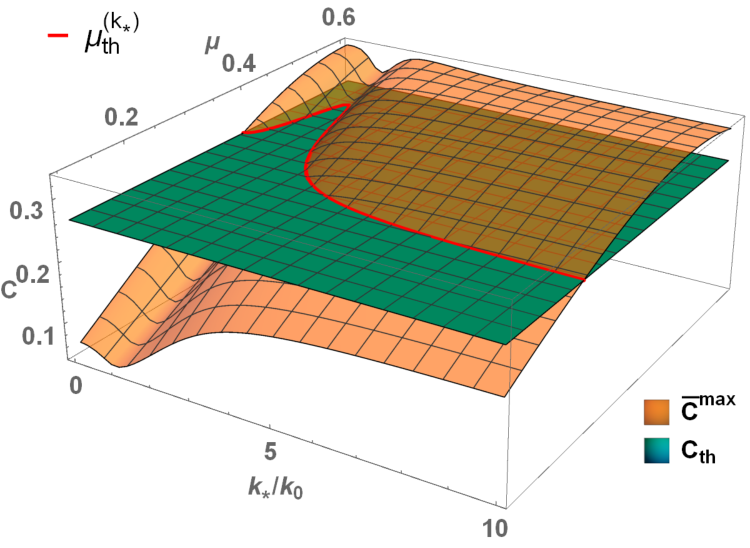}
\hspace{0.03\textwidth}
\includegraphics[width=0.3\textwidth]{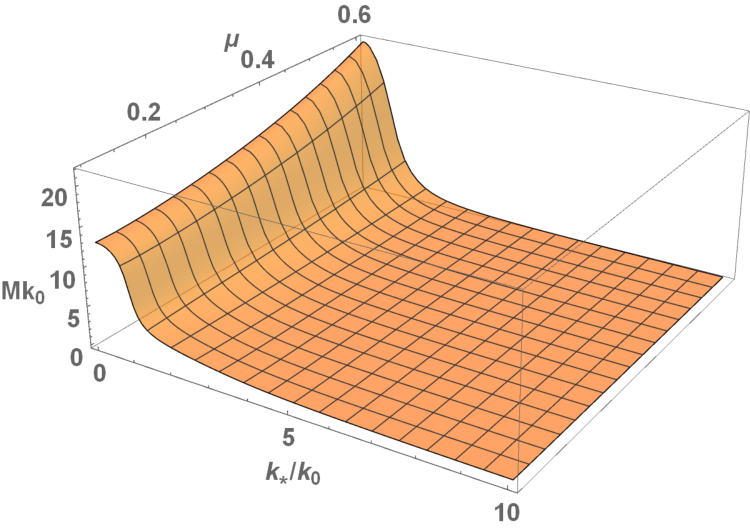}
\end{center}
\caption{(Left panel) $r_{\rm m}(\mu,k_*)$, (middle panel) $\bar \calC^{\rm max}(\mu,k_*)$, $\calC_{\rm th}$ and $\mu^{(k_*)}_{\rm th}$, and (right panel) $\bar M(\mu,k_*)$ for $\sigma=0.1$ and $3f_{\rm NL}/5=1$.}
\label{fig:extended0}
\end{figure}

Let us consider the simple extended power spectrum given by 
\begin{equation}
\calP(k)
=
3 \sqrt{\frac{6}{\pi}} \sigma^2 \left( \frac{k}{k_0} \right)^3
\exp \left( -\frac{3}{2} \frac{k^2}{k_0^2} \right)
\, . 
\label{eq:Gausspower}
\end{equation}
For this case, 
the functional form of $\psi(r)$ is given by 
\begin{equation}
\psi(r) = \exp \left( -\frac{k_0^2r^2}{6} \right)
\, . 
\end{equation}
The dependence of $r_{\rm m}(\mu,k_*)$, $\bar \calC^{\rm max}(\mu,k_*)$ and $\bar M(\mu,k_*)$ on $k_*$ and $\mu$ is shown in Figure~\ref{fig:extended0} for $\sigma=0.1$ and $3f_{\rm NL}/5=1$. 

\begin{figure}[htb]
\begin{center}
\includegraphics[width=0.45\textwidth]{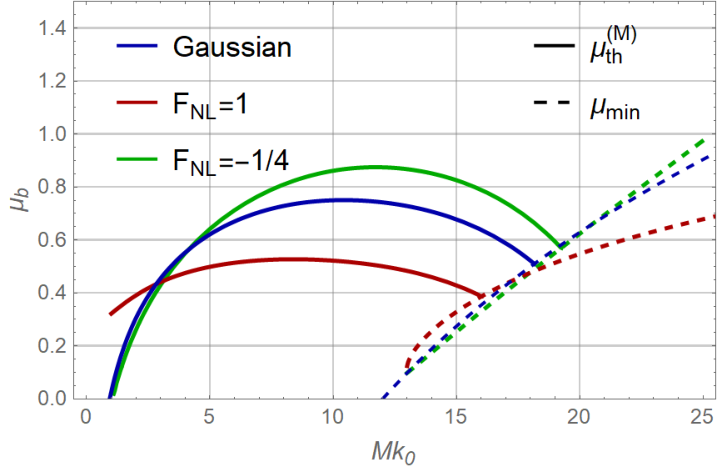}
\hspace{0.03\textwidth}
\includegraphics[width=0.47\textwidth]{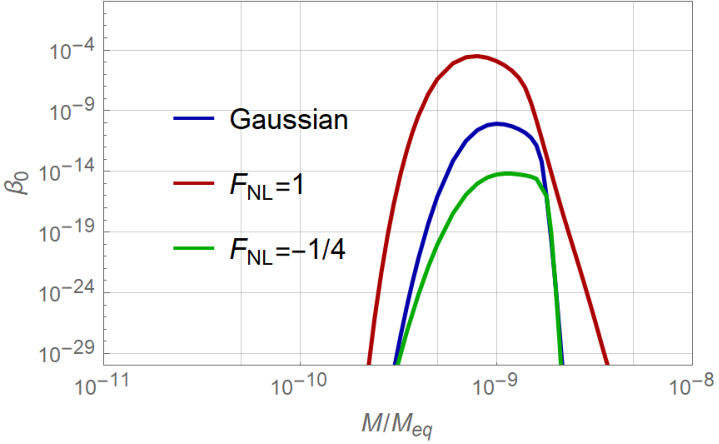}
\end{center}
\caption{
(Left panel) $\mu_{\rm th}(M)$ and $\mu_{\rm min}(M)$ for $\sigma=0.1$, and (right panel) $\beta_0$ as a function of $M$. The PBH abundance increases with a larger value of $f_{\rm NL}$. 
}
\label{fig:extended}
\end{figure}

As we have mentioned, for the extended power spectrum, we find that the maximum value of $M$ is realized at $k_*=0$ for each $\mu$. Therefore, for a given value of $M$, the minimum value of $\mu_{\rm min}$ is given by 
\begin{equation}
\mu_{\rm min}(M) = \mu(M,0)
\, .
\end{equation} 
Thus, in the estimation of the PBH fraction, we should take into account the relevant region of $\mu$ with the mass $M$ given by \eqref{eq:mub}. 
The dependence of $\mu_b(M)$ on different values of $f_{\rm NL}$ is shown in 
the left panel in Figure~\ref{fig:extended}. 
The resultant PBH fraction $\beta_0$ for the extended power spectrum is depicted as a function of $M$ in the right panel of Figure~\ref{fig:extended}.

\section{Summary}
\label{sec:summary}

\begin{figure}[htbp]
\begin{center}
\includegraphics[scale=0.5]{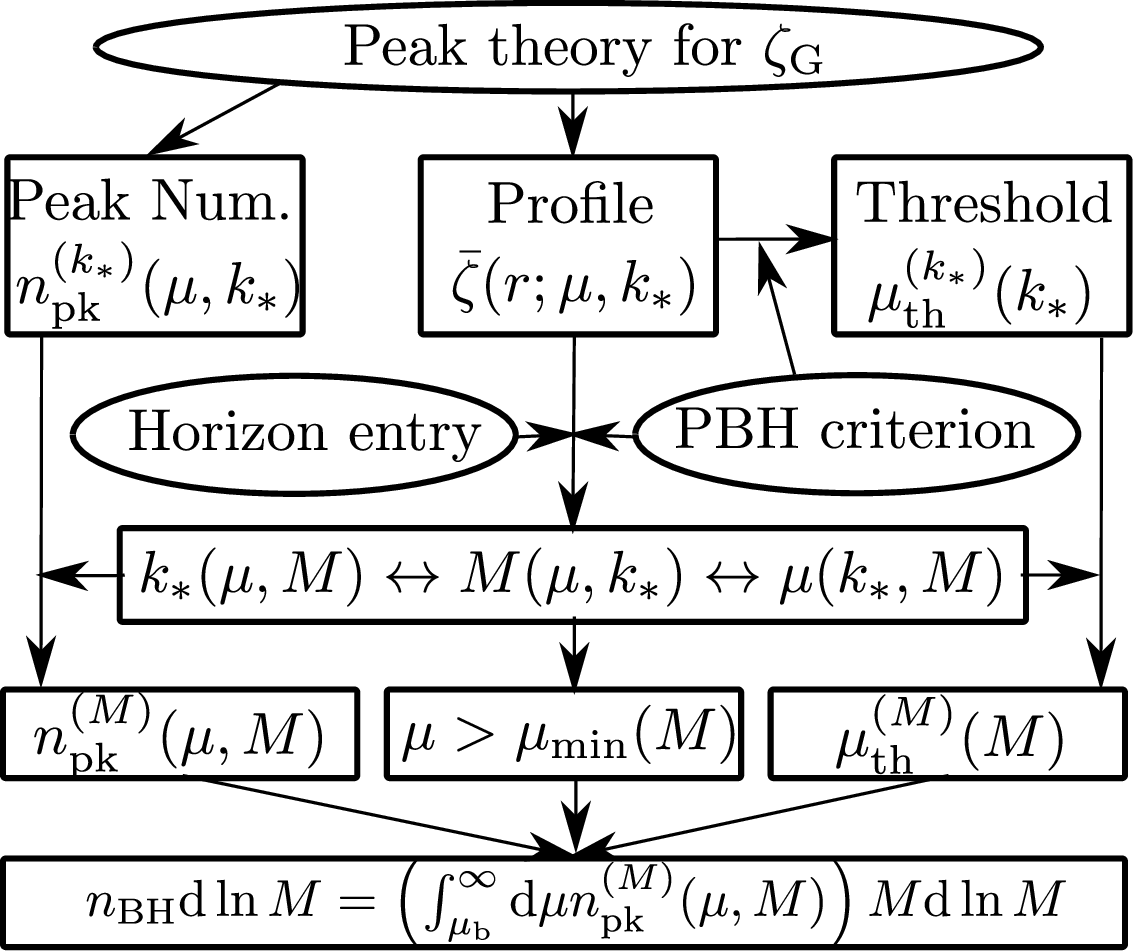}
\caption{A flow chart of our procedure. 
}
\label{fig:flow}
\end{center}
\end{figure}

We have discussed the effect of the local type non-Gaussianity on the PBH abundance. Our procedure is based on the peak theory for the Gaussian variable $\zeta_{\rm G}$, which is related to the non-Gaussian curvature perturbation $\zeta$ via \eqref{eq:nonGzeta}, which is the simplest type of non-Gaussianity with the non-linear parameter $f_{\rm NL}$. Summary of our procedure is schematically shown in Figure~\ref{fig:flow}. The value of $|f_{\rm NL}| \sim 1$ induces a similar effect to a few factors of difference in the amplitude of the power spectrum. The negative/positive value of $|f_{\rm NL}|$ of $\calO(1)$ may reduce/enhance the PBH abundance in several orders of magnitude. The calculation procedure for estimating the PBH abundance shown in this article can be also applied to more general types of primordial non-Gaussianity straightforwardly, once the primordial curvature perturbation is given as a function of the Gaussian variable $\zeta_{\rm G}$.

\noindent
\textit{Note added}: While preparing the manuscript, we found a paper by Atal, Garriga and Marcos-Caballero~\cite{Atal:2019cdz}, which has some overlap with our work.

\subsection*{Acknowledgements}

This work is supported in part by the Japan Society for the Promotion of Science (JSPS) and the National Research Foundation of Korea (NRF) under the Japan-Korea Basic Scientific Cooperation Program (NRF-2018K2A9A2A08000127).
CY is supported in part by JSPS KAKENHI Grant (JP19H01895).
JG is supported in part by the Basic Science Research Program through NRF Research Grant (2016R1D1A1B03930408). 
SY is supported by MEXT KAKENHI Grant Number 15H05888 and 18H04356.

\appendix

\renewcommand{\theequation}{\Alph{section}.\arabic{equation}}

\section{Peaks of $\zeta_{\rm G}$ and $\zeta$}
\label{app:peaks}
\setcounter{equation}{0}

First, let us introduce the following Taylor expansion of the fields $\zeta(x_i)$ an $\zeta_{\rm G}(x_i)$:
\begin{align}
\zeta 
& =
\zeta_0 + \zeta^i_1 x_i + \frac{1}{2} \zeta_2^{ij} x_ix_j + \calO(x^3)
\, , 
\\
\zeta
& =
\zeta_{\rm G0} + \zeta_{\rm G1}^i x_i + \frac{1}{2} \zeta_{\rm G2}^{ij} x_ix_j + \calO(x^3)
\, . 
\end{align}
With \eqref{eq:nonGzeta}, the coefficients satisfy the following relations:
\begin{align}
\zeta_0 
& =
\zeta_{\rm G0} - \frac35 f_{\rm NL} \left( \zeta_{\rm G0}^2 - \sigma_0^2 \right)
\, ,
\label{eq:zeta00}
\\
\zeta_1^i
& = 
\zeta_{\rm G1}^i \left( 1 - \frac65 f_{\rm NL} \zeta_{\rm G0} \right)
\, ,
\label{eq:zeta11}
\\
\zeta_2
& =
\zeta_{\rm G2} \left( 1 - \frac65 f_{\rm NL} \zeta_{\rm G0} \right)
- \frac65 f_{\rm NL} \sum_i \left( \zeta_{\rm G1}^i \right)^2
\, , 
\label{eq:zeta22}
\end{align}
where $\zeta_2=\zeta_2^{11}+\zeta_2^{22}+\zeta_2^{33}$ and $\zeta_{\rm G2}=\zeta_{\rm G2}^{11}+\zeta_{\rm G2}^{22}+\zeta_{\rm G2}^{33}$. Here, we adopt a simple criterion for PBH formation $\delta>\delta_{\rm th}$ at the horizon entry to perform the order-of-magnitude estimation.

The value of $\delta$ can be expressed in terms of the Taylor expansion around a certain point:
\begin{equation}
\delta
=
\frac{4}{9} \frac{1}{a^2H^2} \left\{ e^{2\zeta_0} \left[ \zeta_2 
- \frac{1}{2} \sum_i \left( \zeta_1^i \right)^2 \right] \right\} + \calO(y^2)
\, , 
\label{eq:Taylordelta}
\end{equation}
where we have introduced a new spatial coordinate $\bm y$ to emphasize the difference from the expansion around the peak of $\zeta$. Focusing on the region which will collapse into a PBH, from the inequality $\delta>\delta_{\rm th}$, we obtain the following inequality:
\begin{equation}
\left. \frac{\zeta_2}{\bar a^2 H^2} \right|_\text{horizon entry}
> \frac{9}{4} \delta_{\rm th} \sim 1
\, ,
\label{eq:zeta2ineq}
\end{equation}
where we have defined the renormalized local scale factor $\bar a=ae^{-\zeta_0}$.

Let us apply the inequality \eqref{eq:zeta2ineq} to a peak of $\zeta$. We introduce the following two variables to characterize the profile around a peak of $\zeta$:
\begin{align}
\mu_{\rm NG} & = -\zeta|_{\rm peak}
\, , 
\\
\kappa_{\rm NG} & = \frac{\triangle\zeta|_{\rm peak}}{\mu_{\rm NG}}
\, . 
\end{align}
We note that, while we introduce $k_*^2$ for $\zeta_{\rm G}$ in the main text because we are interested in the case $\triangle \zeta_{\rm G}>0$, $\kappa_{\rm NG}$ can be negative in general. Adopting $\bar a^2 H^2\sim|\kappa|$ as the horizon entry condition, \eqref{eq:zeta2ineq} can be rewritten as 
\begin{equation}
\mu_{\rm NG} > \frac{9}{4} \delta_{\rm th} \sim 1
\, . 
\end{equation}
\eqref{eq:zeta00}, \eqref{eq:zeta11} and \eqref{eq:zeta22} are then written as 
\begin{align}
\mu_{\rm NG}
& = 
\mu + \frac35 f_{\rm NL} \left( \mu^2 - \sigma_0^2 \right)
\, ,
\\
0 & = \zeta_{\rm G1}^i \left( 1 + \frac65 f_{\rm NL} \mu \right)
\, ,
\\
\mu_{\rm NG} \kappa_{\rm NG}
& = 
\mu \kappa \left( 1 + \frac65 f_{\rm NL} \mu \right)
+ \frac65 f_{\rm NL} \sum_i \left( \zeta_{\rm G1}^i \right)^2
\, . 
\label{eq:mukap}
\end{align}
If $1+6f_{\rm NL}\mu/5$ is negative, the sign of $\mu_{\rm NG}\kappa_{\rm NG}$ can be different from $\mu\kappa$. That is, a peak of $\zeta$ can be a minimum of $\zeta_{\rm G}$. In order to avoid this situation, we simply assume $3f_{\rm NL}/5 \gtrsim -1/4$.

\section{PBH formation criteria}
\label{app:criteria}
\setcounter{equation}{0}

Here, we briefly review the estimate on the threshold for the formation of PBHs following~\cite{Harada:2015yda}, introducing the basic perturbation variables used in the main text.

Assuming the spherical symmetry around high peaks, we introduce a compaction function, which represents an
excess of the Misner-Sharp mass $\delta{M}$ in the spherical region with the radius $r$, given by (see~\cite{Harada:2015yda,Yoo:2018esr} for details)
\begin{equation}
\calC(r) 
\equiv
\frac{\delta{M}}{R}
= 
\frac{1}{3} \left[ 1 - \left( 1 - r \zeta' \right)^2 \right]
\, ,
\label{eq:Compaction}
\end{equation}
where $R \equiv are^{-\zeta}$ is the areal radius at the radius $r$, and the second equality holds on the comoving slice. 
Assuming that $\calC$ is a smooth function of $r$, we may find a radius $r=r_{\rm m}$ at which $\calC$ takes the maximum value $\calC^{\rm max} = \calC (r_{\rm m})$, and $\calC'(r_{\rm m}) = 0$ should be satisfied. From \eqref{eq:Compaction}, $\calC'(r_{\rm m}) = 0$ gives the condition for $\zeta$ at $r=r_{\rm m}$ as
\begin{equation}
\left. \left( \zeta' + r \zeta'' \right) \right|_{r=r_{\rm m}} = 0 
\, . 
\label{eq:forrm}
\end{equation}
Then, as a criterion for PBH formation, we consider that PBH is formed when the maximum value of the compaction function is larger than a threshold $\calC_{\rm th}$:
\begin{equation}
\calC^{\rm max}> \calC_{\rm th}
\, . 
\label{maxth}
\end{equation}
In this article, we use $\calC_{\rm th} \approx 0.267$ as a reference value~\cite{Harada:2015yda,Yoo:2018esr}.
The specific value $\calC\approx0.267$ is imported from the results of the numerical simulations for PBH formation 
performed in~\cite{Harada:2015yda}. 
We note that, as is shown in~\cite{Harada:2015yda}, at the moment of the horizon entry $are^{-\zeta}=1/H$, 
the value of the compaction function is equivalent to the half of the volume averaged density perturbation. 
Therefore the inequality \eqref{maxth} can be expressed as $\bar \delta_{\rm max}>\delta_{\rm th}=2\calC_{\rm th}$ 
with $\bar \delta_{\rm max}$ being the maximum value of the averaged density perturbation.

%

Finally, let us give the estimate of the PBH mass. Since the PBH mass is given by $M=\alpha H^{-1}/2$ with $\alpha$ being a numerical factor, from the horizon entry condition $are^{-\zeta} = 1/H$, the PBH mass $M$ can be expressed as follows:
\begin{equation}
M
=
\frac{1}{2}\alpha H^{-1}
=
\frac{1}{2}\alpha a r_{\rm m} e^{-\zeta(r_{\rm m})}
=
M_{\rm eq} k_{\rm eq}^2 r_{\rm m}^2 e^{-2\zeta(r_{\rm m})}
\, , 
\label{eq:kM}
\end{equation}
where we have used $H\propto a^{-2}$ during radiation dominated era, and $M_{\rm eq} = \alpha H_{\rm eq}^{-1}/2$ and $k_{\rm eq} = a_{\rm eq} H_{\rm eq}$ at the matter-radiation equality. The value of the numerical factor $\alpha$ is rather ambiguous, and we set $\alpha=1$ as a fiducial value in the main text.

\end{document}